# Local Cosmological Constant and the Dark Energy Coincidence Problem

**M Tajmar**

Space Propulsion, ARC Seibersdorf research GmbH, A-2444 Seibersdorf, Austria

E-mail: martin.tajmar@arcs.ac.at

**Abstract.** It has been suggested that the Dark Energy Coincidence Problem could be interpreted as a possible link between the cosmological constant and a massive graviton. We show that by using that link and models for the graviton mass a dark energy density can be obtained that is indeed very close to measurements by WMAP. As a consequence of the models, the cosmological constant was found to depend on the density of matter. A brief outline of the cosmological consequences such as the effect on the black hole solution is given.

## 1. Introduction

Recent measurements show that we live in an accelerating and flat universe, which can be interpreted using a small cosmological constant in Einstein's general relativity equations. The origin of this cosmological constant, which is equivalent to a vacuum energy density called dark energy, remains a mystery[1]. Moreover, measurements show that the amount of dark energy accounts for approximately ¾ of the energy content in the universe[2].

Novello[3] recently suggested that this cosmological puzzle can be also interpreted as massive gravitons that are linked to the cosmological constant. There have been some discussions in the literature about a possible link between cosmological constant $\Lambda$ and a massive graviton $m_g$, with some arguing that such a link is not possible[4,5] and with others showing that in an appropriate theoretical framework such a connection can be indeed formulated[6-8].

We extend previous work on that topic and show that the amount of dark energy presently observed in the universe can be indeed calculated using the previously proposed connection between $\Lambda$ and $m_g$. As a consequence of our model, the cosmological constant must then depends on the mass density and hence on the domain which is observed.

## 2. Dark Energy and Cosmological Constant

The Friedman equations including a cosmological constant for a flat expanding universe are known as

$$H^2 = \frac{8\pi G}{3}\rho + \frac{\Lambda c^2}{3}, \qquad (1)$$

where $\rho$ is the homogeneous density source of the universe model and H is the Hubble's function. The dark energy density is commonly defined with respect to the critical density of a flat universe as

$$\Omega_\Lambda = \left(\frac{8\pi G}{3H^2}\right)\rho_\Lambda. \qquad (2)$$

Using $\rho_\Lambda = \frac{c^2}{8\pi G}\Lambda$ and combining Equs. (1) and (2), we can express the dark energy density as

$$\Omega_\Lambda = \frac{1}{1 + \frac{8\pi G}{\Lambda c^2}\rho} \quad . \tag{3}$$

Novello and others[6-8] proposed a link between a cosmological constant and a graviton mass,

$$\frac{1}{\lambda_g^2} = \frac{m_g^2 c^2}{\hbar^2} = \frac{2}{3}\Lambda \quad , \tag{4}$$

where $\lambda_g$ is the graviton wavelength. This follows naturally from linearizing Einstein's field equation for gravity including a cosmological constant as well as from the equations of motion for a massive spin-2 field propagating in a de-Sitter background. If we now find a model for the graviton mass, we can calculate the cosmological constant and finally also the dark energy density $\Omega_\Lambda$. At least two such models were proposed in the literature. Argyris et al[9] obtained the graviton mass by solving Einstein's equations in the conformally flat case and comparing it with Proca-type solutions of a weak field approximation to general relativity. In general, the Proca equations include the effect of a massive exchange particle (in our case a graviton) into Maxwell-type field equations. As a result, the graviton mass arises from the interaction with matter and is given by

$$m_g = \frac{8\hbar}{c^2}\left(\frac{\pi G}{3}\rho\right)^{0.5} \quad . \tag{5}$$

A similar result was recently obtained by the author, showing that the gravitomagnetic Larmor theorem[10] (an observer can not distinguish between a rotating reference frame or a gravitomagnetic field, $B_g=-2\omega$) can be expressed by a local graviton mass and Proca-type solutions of weak gravitational fields as[11]

$$m_g = \frac{2\hbar}{c^2}(\pi G \rho)^{0.5} \quad . \tag{6}$$

The shortcoming of these solutions is that they depend on a Proca model of weak gravitational fields. That leads to a Spin-1 graviton which does not contain the full solutions to general relativity theory. However, on the scale of the universe and even on the local laboratory scale, the weak field approximation to general relativity does apply and the models used should be applicable.

Using Equ. (4) and Equs. (5) and (6), we can express the cosmological constant as a function of the local mass density,

$$\Lambda = \alpha \mu_{0g} \rho \quad , \tag{7}$$

where $\mu_{0g}$ is the gravitomagnetic permeability of free space ($\frac{4\pi G}{c^2}$) and $\alpha=1.5$ in the Tajmar graviton model or $\alpha=8$ in case of the Argyris graviton model. Table 1 lists examples of the cosmological constant computed from Equ. (7) for various representative environments from the scale of our Earth to the universe. Note that the solution of the cosmological constant for the universe fits the predictions from WMAP measurements[2].

We can now solve for the dark energy density in the universe by using Equ. (7) in Equ. (4) and finally get

$$\Omega_\Lambda = \frac{\alpha}{2+\alpha} \ . \tag{8}$$

Due to the fact that both the cosmological constant in our model depends on the mass density as well as the dark energy density, we obtain a simple solution that is density invariant.

**Fig. 1** shows how the dark energy density depends on our pre-factor $\alpha$. For $\alpha=1.5$ (Tajmar model), we get $\Omega_\Lambda=0.43$ and for $\alpha=8$ (Argyris model), we get $\Omega_\Lambda=0.8$, which is already very close to the WMAP measurement of $\Omega_\Lambda=0.73\pm0.04$. That provides a natural explanation for the dark energy coincidence problem, which asks why the density of dark matter is similar to the density of dark matter. The density dependent solution of the cosmological constant and Equ (8) offers an explanation for this coincidence as the dark energy density $\Omega_\Lambda$ between 0.43 – 0.8 is indeed comparable to the observed density of matter $\Omega_M=0.22$.

**3. Discussion**
This is a remarkable result showing that the presently observed dark energy density in the universe can be explained by models for the graviton mass and its relation to the cosmological constant without adjustment of any numerical factors. If these models really apply, what are the cosmological consequences? We are then led to the conclusion that the dark energy density is a result of massive gravitons that depend on the local mass density. As the solution is density invariant, the present dark energy density would then also apply to the early stages of the universe where the density was greater. The solution should always apply independent on the age of the universe. It is a constant pressure within the universe model.

Does it alter any well-known solutions? Of course, the consequences of the interpretations above must be thought through all cosmological scenarios. Apart from the first immediate consequence, that there should be then a dispersion relation for the group velocity of gravity depending on the local mass density $v_g = c \cdot \sqrt{1 - 4\pi G \frac{\rho_m}{\omega^2}}$, we would like to assess probably the most popular cosmological object, a black hole.

The Schwarzschild metric in de-Sitter spacetime can be written as[12]

$$d\tau^2 = \left(1 - \frac{2MG}{c^2 r} - \frac{\Lambda r^2}{3}\right)dt^2 - \left(1 - \frac{2MG}{c^2 r} - \frac{\Lambda r^2}{3}\right)^{-1} dr^2 - r^2\left(d\theta^2 + \sin^2\theta d\phi^2\right) . \tag{9}$$

Using the expression for the cosmological constant in Equ. (7), we can therefore express the Schwarzschild radius using the metric in Equ. (9) as

$$r_S = \frac{MG}{c^2}(2+\alpha) \ . \tag{10}$$

This deviates from the classical result by a factor of $\frac{2}{2+\alpha}$, which is small for our $\alpha$ used. However, there is an important difference with respect to previous work on this subject. Previously it was thought that the gravitational horizon $L=\Lambda^{-0.5}$ is always much larger than the Schwarzschild radius, because a cosmological constant of $1.29 \times 10^{-52}$ m$^2$ was assumed. In our case, the cosmological constant of a black hole is given by

$$\Lambda_S = \frac{c^4}{M^2 G^2} \frac{3\alpha}{(2+\alpha)^3} \ . \tag{11}$$

This gives a gravitational horizon of

$$L_S = \frac{1}{\sqrt{\Lambda_S}} = \frac{MG}{c^2}\left[\frac{(2+\alpha)^3}{3\alpha}\right]^{0.5}. \qquad (12)$$

For α smaller than 1, the ratio of $\frac{L_S}{\Lambda_S}$ is approaching infinity, however, for α greater than 1, the ratio approaches $\frac{1}{\sqrt{3}}$. For our two graviton mass solutions, $\frac{L_S}{\Lambda_S}$ gives 0.88 and 0.65 respectively. This has the consequence, that the gravitational force outside the black hole should now decrease with a Yukawa type modification affecting gravitational fields (that is why the Schwarzschild radius is larger compared to the classical solution). That can be an important input to solve presently observed anomalies such as nested discs in rotating Keplerian rotation around supermassive black holes[13].

**4. Conclusion**
Our analysis strengthens the suggestion that the dark energy puzzle in cosmology can be interpreted as a consequence of a link between a graviton mass and the cosmological constant. We have shown that present models for the graviton mass and the link to the cosmological constant indeed leads to a dark energy density which is very close to the one that is presently observed. As a consequence, the graviton mass and the cosmological constant should depend on the mass density. The dark energy density in our models was however found to be invariant with mass density. It would then appear that the presently observed dark energy density is independent on the age of the universe and homogenously distributed throughout the universe.

**Table 1.** Cosmological Constant Examples (for $\alpha=1.5$)

| Location | Cosmological Constant [m²] |
|---|---|
| Sun | $1.97 \times 10^{-23}$ |
| Earth | $7.68 \times 10^{-23}$ |
| Solar System | $3.14 \times 10^{-35}$ |
| Milky Way | $6.29 \times 10^{-48}$ |
| Universe | $1.29 \times 10^{-52}$ |

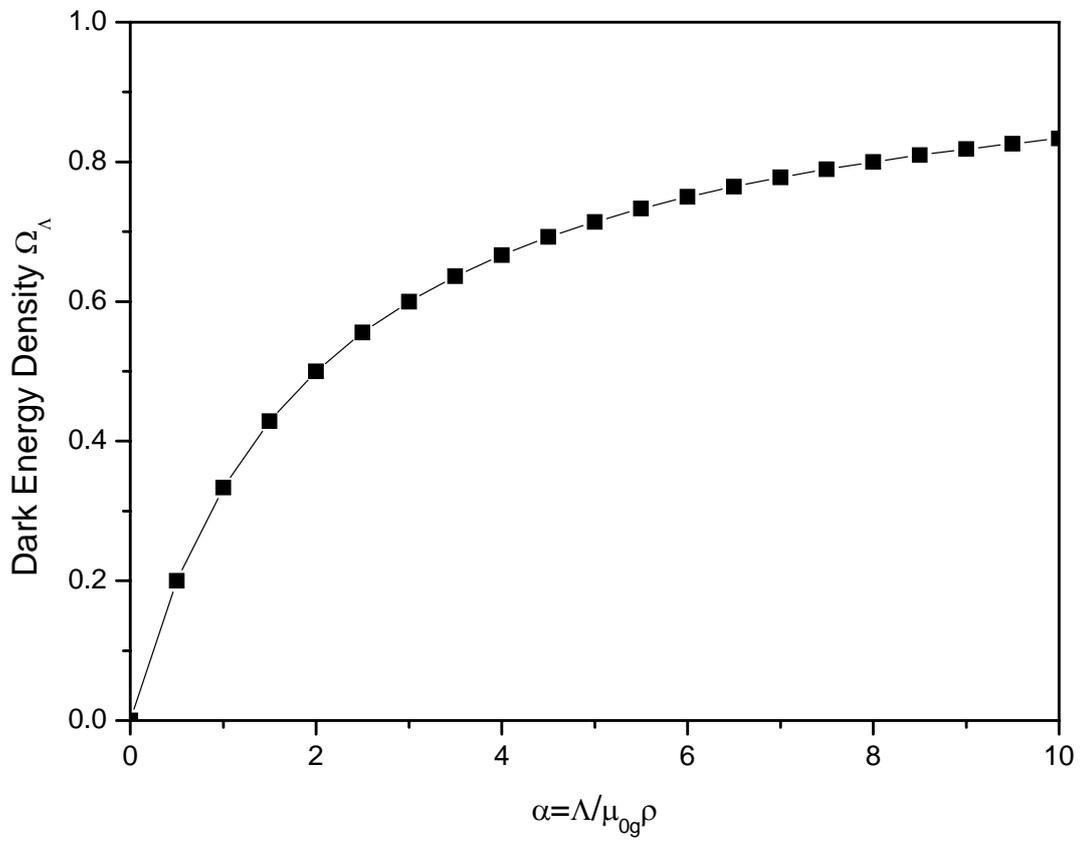

**Figure 1.** Variation of Dark Energy Density on Pre-Factor α